\documentclass[onecolumn,preprintnumbers,amsmath,aps]{revtex4}
\usepackage{graphicx}
\usepackage{dcolumn}
\usepackage{bm}
\usepackage{subfigure}
\usepackage{color}
\begin{document}
%%%%%%%%%%%%%%%%%%%%%%%%%%%%
\def\eq#1{\ref{#1}}
\def\fig#1{\ref{#1}}
\def\tab#1{\ref{#1}}
%%%%%%%%%%%%%%%%%%%%%%%%%%%%
\title{Description of the superconducting state in the high-pressure fcc phase of platinum-hydride}
\author{R. Szcz{\c{e}}{\'s}niak$^{\left(1\right)}$, D. Szcz{\c{e}}{\'s}niak$^{\left(2, 3\right)}$, K.M. Huras$^{\left(1\right)}$}
%%%%%%%%%%%%
\affiliation{1. Institute of Physics, Cz{\c{e}}stochowa University of Technology, Al. Armii Krajowej 19, 42-200 Cz{\c{e}}stochowa, Poland}
%%%%%%%%%%%%
\email{khuras@wip.pcz.pl}
%%%%%%%%%%%%
\affiliation{2. Institute of Physics, Jan D{\l}ugosz University in Cz{\c{e}}stochowa, Al. Armii Krajowej 13/15, 42-200 Cz{\c{e}}stochowa, Poland}
\affiliation{3. Institute for Molecules and Materials UMR 6283, University of Maine, Ave. Olivier Messiaen, 72085 Le Mans, France}
%%%%%%%%%%%%
\date{\today} 
\begin{abstract}
%%%%%%%%%%%%%%%%%%%%%%%%%%%%%%%%%%%%%%%%%%%%%%%%%%%
The thermodynamic parameters of the superconducting state in PtH at the pressure $76$ GPa have been examined. The calculations have been carried out in the framework of the Eliashberg formalism. It has been found that the critical temperature ($T_{C}$) changes in the range from $30.6$ K to $16.8$ K, depending on the assumed value of the Coulomb pseudopotential: $\mu^{\star}\in\left<0.1,0.3\right>$. The other thermodynamic quantities differ significantly from the predictions of the classical BCS theory.
In particular: (i)  The parameter $2\Delta\left(0\right)/k_{B}T_{C}$ reaches the values from $4.25$ to $3.98$, where $\Delta\left(0\right)$ denotes the low-temperature order parameter.
(ii) The ratio of the specific heat jump ($\Delta C$) to the specific heat in the normal state ($C^{N}$) takes the values from $2.07$ to $1.97$.
(iii) Finally, $T_{C}C^{N}\left(T_{C}\right)/H^{2}_{C}\left(0\right)\in\left<0.145, 0.156\right>$, where $H_{C}\left(0\right)$ is the low-temperature thermodynamic critical field.
%%%%%%%%%%%%%%%%%%%%%%%%%%%%%%%%%%%%%%%%%%%%%%%%%%%
\end{abstract}
\maketitle

\section{Introduction}
The electron-phonon interaction can lead to the formation of the superconducting state that is characterized by a high value of the critical temperature ($T_{C}$). In the first case, it is relevant for the multi-band systems, in the second case, for the high-temperature superconductors (cuprates), and in the third case, for the materials under the influence of high pressure ($p$). 

An example of the multi-band system, in which the phonon-mediated superconducting state is induced at a high value of the critical temperature, is magnesium diboride \cite{Nagamatsu}. On the basis of the conducted experiments, it has been stated that $\left[T_{C}\right]_{\rm MgB_{2}}$ equals $39.4$ K. From the physical point of view, the anomalous thermodynamic properties of the considered superconducting state are related to the existence of two energy gaps; the first one value $(7.0-7.4)$ meV corresponds to the two-dimensional $\sigma$ band, the second one $(2.6-2.8)$ meV exists in the three-dimensional $\pi$ band \cite{Radek01}, \cite{Moca}.

In the case of cuprates, some authors have assumed that the electron-phonon interaction, supplemented by the electron-electron-phonon interaction, can be responsible for the creation of the superconducting phase \cite{Radek02}, \cite{Kim2}, \cite{KulicA}, \cite{KulicB}. Note that cuprates are the materials characterized by the highest experimentally verified values of $T_{C}$. For example, for the compound ${\rm HgBa_{2}Ca_{2}Cu_{3}O_{8+y}}$ (Hg1223), the following has been obtained $T_{C}\simeq 135$ K \cite{Chu}, wherein, under the influence of the pressure ($p\simeq 31$ GPa), the critical temperature in Hg1223 increases to the value of about $164$ K \cite{GaoHTSC}. 

As it turns out, the pressure may be the factor that causes the strong increase of the electron-phonon interaction \cite{Schilling}. In particular, it is clearly visible in the case of lithium and calcium, where the maximum value of $T_{C}$ is equal to $14$ K ($p=30.2$ GPa) and to $25$ K ($p=161$ GPa), respectively \cite{Deemyad}, \cite{Yabuuchi}. 

Additionally, it is believed that a very high value of the critical temperature can be obtained in metallic hydrogen \cite{Ashcroft}. For instance, for the pressure at $450$ GPa, the superconducting state in the molecular phase of hydrogen is characterized by $T_{C}\simeq 240$ K \cite{Cudazzo}. At the extremely high pressure ($p\sim 2$ TPa), the critical temperature may be even equal to $600$ K \cite{Maksimov}, \cite{Radek03}. It is worth noting that from the physical point of view, the high value of the critical temperature in hydrogen is associated with the small mass of the atomic nuclei forming the crystal lattice and the lack of the inner electron shells, which results in the high value of the electron-phonon coupling constant. 

In order to obtain the superconducting state with a high value of the critical temperature, but at the pressure that is significantly lower than the metallization pressure for hydrogen ($p\sim 400$ GPa), Ashcroft proposed the method of the chemical pre-compression \cite{AshcroftWodorowane}. The suggestion made by Ashcroft caused a rapid increase in the number of the publications on this topic. In particular, the following chemical compounds have been taken into consideration: ${\rm CH_{4}}$, ${\rm GeH_{4}}$, ${\rm SnH_{4}}$, and ${\rm SiH_{4}}$  \cite{Tse}, \cite{Gao}, \cite{Canales}, \cite{Gao1}, \cite{Chen}, \cite{Eremets}. We can notice that the obtained theoretical results have been so promising that this branch of physics is now one of the most intensively developed. 

In the presented paper, we have discussed the results obtained for PtH under the pressure at $76$ GPa. In particular, we have determined the most important thermodynamic parameters of the superconducting state in the framework of the Eliashberg formalism \cite{Eliashberg}, \cite{Kim}.     

%%%%%%%%%%%%%%%%%%%%%%%%%%%%%%%%%%%%%%%%%%%%%%%%%%%
\section{The Eliashberg formalism on the imaginary axis}
%%%%%%%%%%%%%%%%%%%%%%%%%%%%%%%%%%%%%%%%%%%%%%%%%%%

%%%%%%%%%%%%%%%%%%%%%%%%%%%%%%%%%%%%%%%%%%%%%%%%%%%
\subsection{The Eliashberg equations}
%%%%%%%%%%%%%%%%%%%%%%%%%%%%%%%%%%%%%%%%%%%%%%%%%%%

The Eliashberg formalism on the imaginary axis ($i\equiv\sqrt{-1}$) allows us to determine the values of the order parameter function ($\phi_{n}\equiv\phi\left(i\omega_{n}\right)$) and the wave function renormalization factor ($Z_{n}\equiv Z\left(i\omega_{n}\right)$). The symbol $\omega_{n}$ denotes the $n$-th Matsubara frequency: $\omega_{n}\equiv \left(\pi / \beta\right)\left(2n-1\right)$, where $\beta\equiv\left(k_{B}T\right)^{-1}$ and $k_{B}$ is the Boltzmann constant. We notice that the order parameter is defined by the ratio: $\Delta_{n}\equiv \phi_{n}/Z_{n}$.

The functions $\phi_{n}$ and $Z_{n}$ can be calculated using the equations below \cite{Carbotte01}, \cite{Carbotte02}, \cite{Allen}:
\begin{equation}
\label{r1}
\phi_{n}=\frac{\pi}{\beta}\sum_{m=-M}^{M}
\frac{\lambda\left(i\omega_{n}-i\omega_{m}\right)-\mu^{\star}\theta\left(\omega_{c}-|\omega_{m}|\right)}
{\sqrt{\omega_m^2Z^{2}_{m}+\phi^{2}_{m}}}\phi_{m},
\end{equation}
\begin{equation}
\label{r2}
Z_{n}=1+\frac{1}{\omega_{n}}\frac{\pi}{\beta}\sum_{m=-M}^{M}
\frac{\lambda\left(i\omega_{n}-i\omega_{m}\right)}{\sqrt{\omega_m^2Z^{2}_{m}+\phi^{2}_{m}}}
\omega_{m}Z_{m},
\end{equation}
where the pairing kernel for the electron-phonon interaction is given by the formula: 
\begin{equation}
\label{r3}
\lambda\left(z\right)\equiv 2\int_0^{\Omega_{\rm{max}}}d\Omega\frac{\Omega}{\Omega ^2-z^{2}}\alpha^{2}F\left(\Omega\right).
\end{equation}

In equation (\ref{r3}), the $\alpha^{2}F\left(\Omega\right)$ denotes the Eliashberg function. For the PtH compound at 76 GPa, this function has been calculated in the paper \cite{Kim}. The value of the maximum phonon frequency ($\Omega_{\rm{max}}$) is equal to $117.81$ meV.

The electron correlations have been modeled by the Coulomb pseudopotential $\mu^{\star}$; $\theta$ denotes the Heaviside unit function and $\omega_{c}$ is the cut-off frequency; $\omega_{c}=3\Omega_{\rm{max}}$. This study refers to a wide range of the Coulomb pseudopotential: $\mu^{\star}\in\left<0.1,0.3\right>$. 

The Eliashberg equations on the imaginary axis compose the infinite system of the equations. However, for the temperatures greater than the temperature of zero Kelvin one can reduce their number due to the fact that $\phi_{n}$ and $Z_{n}$ become saturated at large values of $n$. In particular, the following range of the temperature has been assumed: $T\in\left<T_{0}=2.5{\rm K},T_{C}\right>$. In the present case, $2201$ Matsubara frequencies ($M=1100$) should be taken into consideration.

The Eliashberg equations have been solved using the iterative method, presented and tested in the papers \cite{Radek04} and \cite{Radek05}.

%%%%%%%%%%%%%%%%%%%%%%%%%%%%%%%%%%%%%%%%%%%%%%%%%%%
\subsection{The order parameter and the wave function renormalization factor}
%%%%%%%%%%%%%%%%%%%%%%%%%%%%%%%%%%%%%%%%%%%%%%%%%%%

\begin{figure}[th]
\includegraphics[scale=0.25]{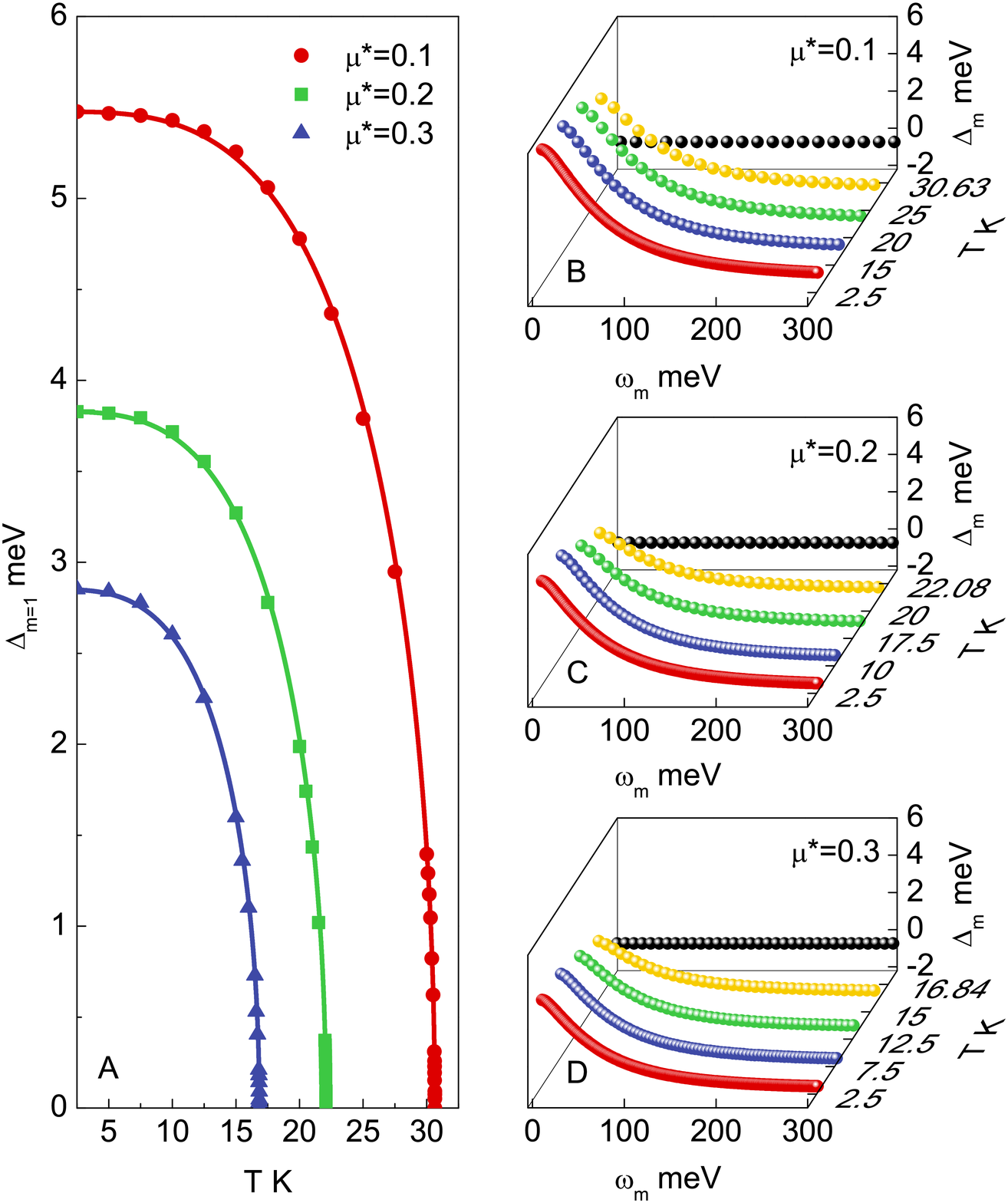}
\caption{
(A) The order parameter $\Delta_{m=1}$ as a function of the temperature for the selected values of the Coulomb pseudopotential. 
The circles, squares and triangles correspond to the exact numerical solutions of the Eliashberg equations. The solid lines represent the results obtained by 
using analytical scheme (see Eq.(\ref{r4})). (B)-(D) The order parameter on the imaginary axis for the selected values of $T$ and $\mu^{\star}$.}
\label{f1}
\end{figure}

Figure \ref{f1} (A) presents the dependence of the order parameter $\Delta_{m=1}$ on the temperature and the Coulomb pseudopotential. The charts have been drawn up on the basis of the results presented in Figs. \ref{f1} (B)-(D).  

%
%%%%%%%%%%%%%%%%%%%%%%%%%%%(Rysunek.1.)
%%%%%%%%%%%%%%%%%%%%%%%%%%%
%

It can be noted that the value of the function $\Delta_{m=1}\left(T,\mu^{\star}\right)$ strongly decreases with the increasing temperature and the Coulomb pseudopotential. As it turns out, the course of the order parameter can be parameterized using the simple formula:
\begin{equation}
\label{r4}
\Delta_{m=1}\left(T,\mu^{\star}\right)=\Delta_{m=1}\left(\mu^{\star}\right)\sqrt{1-\left(\frac{T}{T_{C}}\right)^{\Gamma}}, 
\end{equation}
where $\Gamma=3.35$, and:
\begin{equation}
\label{r5}
\Delta_{m=1}\left(\mu^{\star}\right)=33.57\left(\mu^{\star}\right)^{2}-26.56\mu^{\star}+7.80.
\end{equation}
It should be noted that the critical temperature is also strongly dependent on $\mu^{\star}$, and it belongs to the range from $30.6$ K to $16.8$ K.

From the physical point of view, the quantity of $2\Delta_{m=1}\left(T,\mu^{\star}\right)$ with a good accuracy determines the value of the energy gap at the Fermi level. Thus, the formula (\ref{r5}) is relatively accurate to estimate the physical value of the Coulomb pseudopotential on the basis of the tunnel experiment \cite{McMillanRowell}.   

\begin{figure}[th]
\includegraphics[scale=0.25]{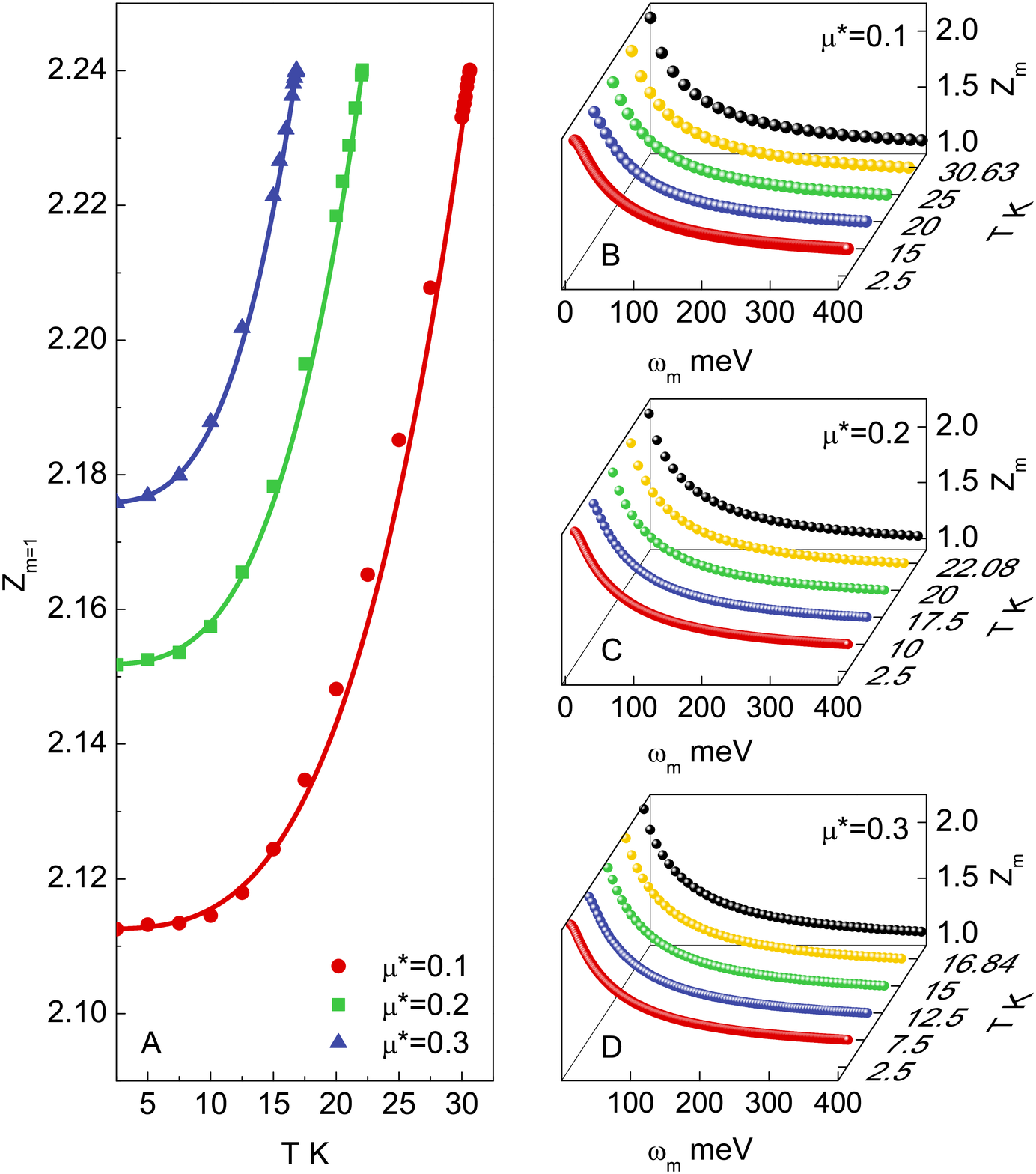}
\caption{
(A) The wave function renormalization factor $Z_{m=1}$ as a function of the temperature for the selected values of the Coulomb pseudopotential. The circles, squares and triangles correspond to the exact numerical solutions of the Eliashberg equations. The solid lines represent the results obtained by using analytical scheme (see Eq.(\ref{r6})).
(B)-(D) The renormalization factor on the imaginary axis for the selected values of $T$ and $\mu^{\star}$.}
\label{f2}
\end{figure}

Figure \ref{f2} (A) presents the dependence of the wave function renormalization factor ($Z_{m=1}$) on the temperature and the selected values of the Coulomb pseudopotential. These charts are based on the data collected in Figs. \ref{f2} (B)-(D).

%
%%%%%%%%%%%%%%%%%%%%%%%%%%%(Rysunek.2.)
%%%%%%%%%%%%%%%%%%%%%%%%%%%
%

In contrast to the order parameter, the function $Z_{m=1}\left(T,\mu^{\star}\right)$ grows together with the increase of $T$ and $\mu^{\star}$. However, the changes in the value of $Z_{m=1}\left(T,\mu^{\star}\right)$ are not too large. 

It is worth noting that the wave function renormalization factor for the first Matsubara frequency can be parameterized with the use of the following expression:
\begin{eqnarray}
\label{r6}
\nonumber
Z_{m=1}\left(T,\mu^{\star}\right)&=&\left[Z_{m=1}\left(T_{C}\right)-Z_{m=1}\left(\mu^{\star}\right)\right]\left(\frac{T}{T_{C}}\right)^{\Gamma}\\
&+&Z_{m=1}\left(\mu^{\star}\right).
\end{eqnarray}
The value of $Z_{m=1}\left(T_{C}\right)$ can be calculated on the basis of the formula: $Z_{m=1}\left(T_{C}\right)=1+\lambda$, where $\lambda$ denotes the electron-phonon coupling constant:
${\lambda\equiv 2\int^{\Omega_{\rm{max}}}_0 \alpha^2\left(\Omega\right)F\left(\Omega\right)/\Omega}$. For the PtH compound, we have obtained: $\lambda=1.24$. The function $Z_{m=1}\left(\mu^{\star}\right)$ has the form:
$Z_{m=1}\left(\mu^{\star}\right)=-0.764\left(\mu^{\star}\right)^{2}+0.622\mu^{\star}+2.058$.

From the physical point of view, the value of $Z_{m=1}\left(T,\mu^{\star}\right)$ with a good approximation determines the dependence of the electron effective mass ($m^{\star}_{e}$) on the temperature and the Coulomb pseudopotential: $m^{\star}_{e}=Z_{m=1}\left(T,\mu^{\star}\right)m_{e}$, where the symbol $m_{e}$ represents the electron band mass. Based on the results presented in figure \ref{f2} (A), it can be noticed that the effective mass of the electron in the PtH compound is high in the entire range of the considered temperatures. 

%%%%%%%%%%%%%%%%%%%%%%%%%%%%%%%%%%%%%%%%%%%%%%%%%%%
\subsection{The thermodynamic critical field and the specific heat}
%%%%%%%%%%%%%%%%%%%%%%%%%%%%%%%%%%%%%%%%%%%%%%%%%%%

%
%%%%%%%%%%%%%%%%%%%%%%%%%%%(Rysunek.3.)
\begin{figure}[th]
\includegraphics[scale=0.25]{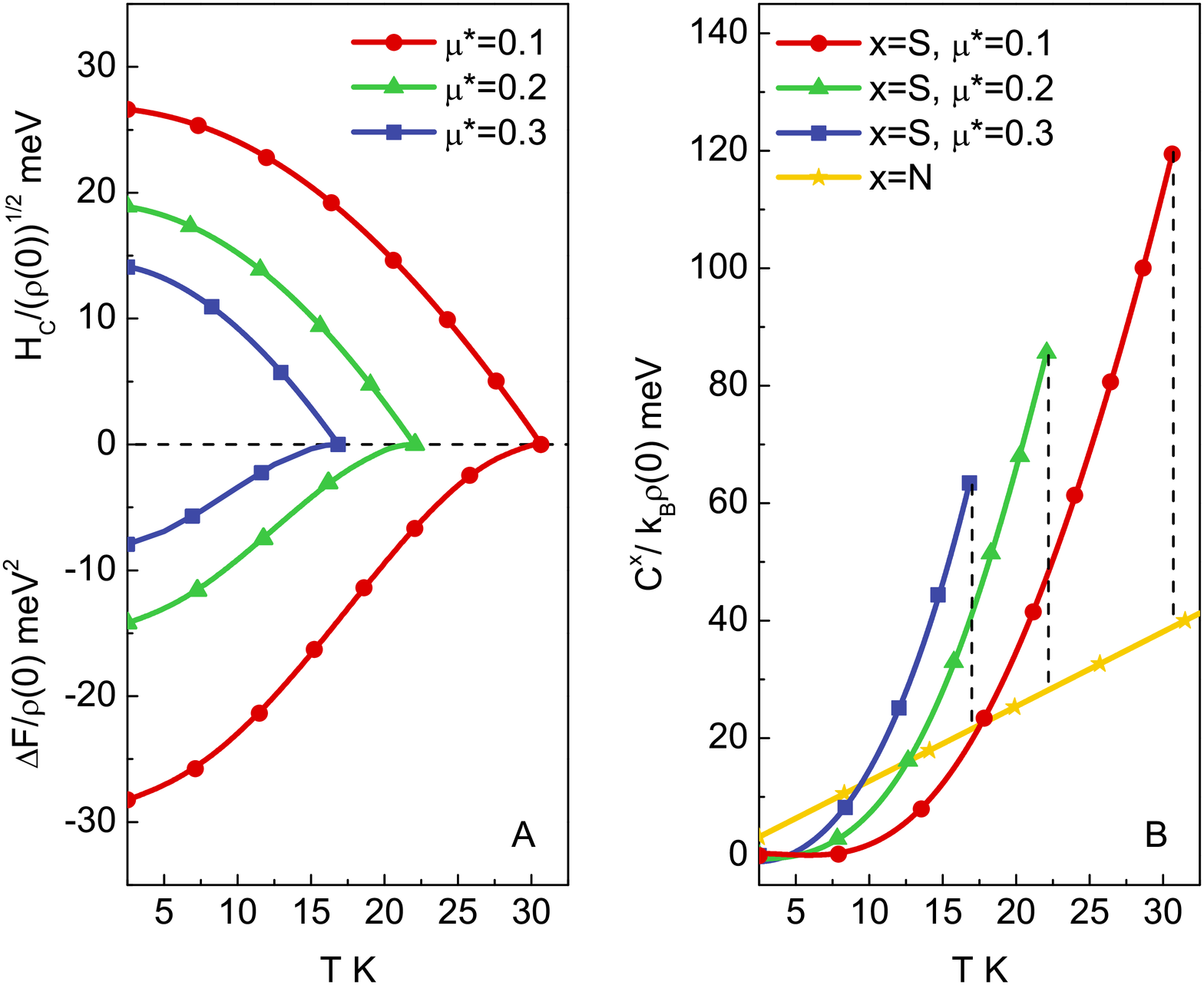}
\caption{
(A) The difference in the free energy (lower panel) and the thermodynamic critical field (upper panel) as a function of the temperature for the selected values of the Coulomb pseudopotential. 
(B) The specific heat of the superconducting and normal state as a function of the temperature for the selected values of the Coulomb pseudopotential.}
\label{f3}
\end{figure}
%%%%%%%%%%%%%%%%%%%%%%%%%%%
%

The thermodynamic critical field and the specific heat can be determined by calculating the difference in the free energy between the superconducting and the normal state \cite{Bardeen}: 
\begin{eqnarray}
\label{r7}
\frac{\Delta F}{\rho\left(0\right)}&=&-\frac{2\pi}{\beta}\sum_{n=1}^{M}
\left(\sqrt{\omega^{2}_{n}+\Delta^{2}_{n}}- \left|\omega_{n}\right|\right)\\ \nonumber
&\times&(Z^{S}_{n}-Z^{N}_{n}\frac{\left|\omega_{n}\right|}
{\sqrt{\omega^{2}_{n}+\Delta^{2}_{n}}}),
\end{eqnarray}  
where $\rho\left(0\right)$ is the value of the electron density of states at the Fermi level. The symbols $Z^{S}_{n}$ and $Z^{N}_{n}$ denote the wave function renormalization factor for the superconducting state ($S$) and the normal state ($N$), respectively.

In the lower panel of figure \ref{f3} (A), we have plotted the dependence of $\Delta F/\rho\left(0\right)$ on the temperature for the selected values of the Coulomb pseudopotential. It has been found that with the increase of the parameter $\mu^{\star}$, the absolute values of the difference in the free energy strongly decrease. The discussed trend can be characterized by the ratio: 
$\left[\Delta F\left(T_{0}\right)\right]_{\mu^{\star}=0.1}/\left[\Delta F\left(T_{0}\right)\right]_{\mu^{\star}=0.3}=3.55$.
     
The next step is to determine the thermodynamic critical field:
\begin{equation}
\label{r8}
\frac{H_{C}}{\sqrt{\rho\left(0\right)}}=\sqrt{-8\pi\left[\Delta F/\rho\left(0\right)\right]}.
\end{equation}

The upper panel of the figure \ref{f3} (A) shows the obtained results. Similarly as it has been observed for the free energy, the increase in the Coulomb pseudopotential causes a severe decline in the value of the thermodynamic critical field: 
$\left[H_{C}\left(0\right)\right]_{\mu^{\star}=0.1}/\left[H_{C}\left(0\right)\right]_{\mu^{\star}=0.3}=1.89$, where $H_{C}\left(0\right)\equiv H_{C}\left(T_{0}\right)$.

As the next step, the specific heat of the normal state has been estimated: $\frac{C^{N}\left(T\right)}{ k_{B}\rho\left(0\right)}=\frac{\gamma}{\beta}$, where the Sommerfeld constant is given by: $\gamma\equiv\frac{2}{3}\pi^{2}\left(1+\lambda\right)$. On the other hand, the specific heat of the superconducting state should be calculated using the formula: $C^{S}=C^{N}+\Delta C$, where the difference between the specific heat of the superconducting and the normal state ($\Delta C$) equals:
\begin{equation}
\label{r9}
\frac{\Delta C\left(T\right)}{k_{B}\rho\left(0\right)}=-\frac{1}{\beta}\frac{d^{2}\left[\Delta F/\rho\left(0\right)\right]}{d\left(k_{B}T\right)^{2}}.
\end{equation}

The obtained results have been presented in figure \ref{f3} (B). On that basis, we have found the destructive effects of the depairing electron correlations on the value of the specific heat jump at the critical temperature: 
$\left[\Delta C\left(T_{C}\right)\right]_{\mu^{\star}=0.1}/\left[\Delta C\left(T_{C}\right)\right]_{\mu^{\star}=0.3}=1.88$. 

The high value of the electron-phonon coupling constant in the PtH compound means that the values of the determined thermodynamic functions will differ from the expectations of the BCS theory \cite{BCS}. The most convenient way to estimate the differences between the results of the Eliashberg theory and the results obtained in the framework of the BCS model is to calculate the dimensionless ratios:

\begin{equation}
\label{r10}
R_{H}\equiv\frac{T_{C}C^{N}\left(T_{C}\right)}{H_{C}^{2}\left(0\right)},
\quad {\rm and} \quad
R_{C}\equiv\frac{\Delta C\left(T_{C}\right)}{C^{N}\left(T_{C}\right)}.
\end{equation}

The BCS model predicts that the values of $R_{H}$ and $R_{C}$ are universal constants for all the superconductors 
and they are equal to $0.168$ and $1.43$, respectively \cite{BCS}. In the case of the PtH compound, the following results have been obtained: $R_{H}\in\left<0.145, 0.156\right>$ and $R_{C}\in\left<2.07, 1.97\right>$, for $\mu^{\star}\in\left<0.1, 0.3\right>$. On that basis, it is clear that the superconducting state in the PtH compound cannot be correctly described by the BCS model, regardless of the physical value of the parameter $\mu^{\star}$.

%%%%%%%%%%%%%%%%%%%%%%%%%%%%%%%%%%%%%%%%%%%%%%%%%%%
\section{The Eliashberg formalism on the real axis}
%%%%%%%%%%%%%%%%%%%%%%%%%%%%%%%%%%%%%%%%%%%%%%%%%%%

  %%%%%%%%%%%%%%%%%%%%%%%%%%%%%%%%%%%%%%%%%%%%%%%%%%%
\subsection{The Eliashberg equations}
%%%%%%%%%%%%%%%%%%%%%%%%%%%%%%%%%%%%%%%%%%%%%%%%%%%

In the framework of the Eliashberg formalism the precise value of the energy gap at the Fermi level can be calculated on the basis of the course of the order parameter on the real axis ($\omega$). In the examineded case, the form of the function $\Delta\left(\omega\right)$ should be determined using the Eliashberg equation in the mixed representation \cite{MarsiglioMieszana}:  

%
%%%%%%%%%%%%%%%%%%%%%%%%%%%%%%%%%%%%%%%%%%%%%%%%%%%%%%%%%%
%
\begin{eqnarray}
\label{r11}
\phi\left(\omega+i\delta\right)&=&
                                  \frac{\pi}{\beta}\sum_{m=-M}^{M}
                                  \left[\lambda\left(\omega-i\omega_{m}\right)-\mu^{\star}\theta\left(\omega_{c}-|\omega_{m}|\right)\right]
                                  \frac{\phi_{m}}
                                  {\sqrt{\omega_m^2Z^{2}_{m}+\phi^{2}_{m}}}\\ \nonumber
                              &+& i\pi\int_{0}^{+\infty}d\omega^{'}\alpha^{2}F\left(\omega^{'}\right)
                                  \left[\left[N\left(\omega^{'}\right)+f\left(\omega^{'}-\omega\right)\right]
                                  \frac{\phi\left(\omega-\omega^{'}+i\delta\right)}
                                  {\sqrt{\left(\omega-\omega^{'}\right)^{2}Z^{2}\left(\omega-\omega^{'}+i\delta\right)
                                  -\phi^{2}\left(\omega-\omega^{'}+i\delta\right)}}\right]\\ \nonumber
                              &+& i\pi\int_{0}^{+\infty}d\omega^{'}\alpha^{2}F\left(\omega^{'}\right)
                                  \left[\left[N\left(\omega^{'}\right)+f\left(\omega^{'}+\omega\right)\right]
                                  \frac{\phi\left(\omega+\omega^{'}+i\delta\right)}
                                  {\sqrt{\left(\omega+\omega^{'}\right)^{2}Z^{2}\left(\omega+\omega^{'}+i\delta\right)
                                  -\phi^{2}\left(\omega+\omega^{'}+i\delta\right)}}\right],
\end{eqnarray}
and
\begin{eqnarray}
\label{r12}
Z\left(\omega+i\delta\right)&=&
                                  1+\frac{i}{\omega}\frac{\pi}{\beta}\sum_{m=-M}^{M}
                                  \lambda\left(\omega-i\omega_{m}\right)
                                  \frac{\omega_{m}Z_{m}}
                                  {\sqrt{\omega_m^2Z^{2}_{m}+\phi^{2}_{m}}}\\ \nonumber
                              &+&\frac{i\pi}{\omega}\int_{0}^{+\infty}d\omega^{'}\alpha^{2}F\left(\omega^{'}\right)
                                  \left[\left[N\left(\omega^{'}\right)+f\left(\omega^{'}-\omega\right)\right]
                                  \frac{\left(\omega-\omega^{'}\right)Z\left(\omega-\omega^{'}+i\delta\right)}
                                  {\sqrt{\left(\omega-\omega^{'}\right)^{2}Z^{2}\left(\omega-\omega^{'}+i\delta\right)
                                  -\phi^{2}\left(\omega-\omega^{'}+i\delta\right)}}\right]\\ \nonumber
                              &+&\frac{i\pi}{\omega}\int_{0}^{+\infty}d\omega^{'}\alpha^{2}F\left(\omega^{'}\right)
                                  \left[\left[N\left(\omega^{'}\right)+f\left(\omega^{'}+\omega\right)\right]
                                  \frac{\left(\omega+\omega^{'}\right)Z\left(\omega+\omega^{'}+i\delta\right)}
                                  {\sqrt{\left(\omega+\omega^{'}\right)^{2}Z^{2}\left(\omega+\omega^{'}+i\delta\right)
                                  -\phi^{2}\left(\omega+\omega^{'}+i\delta\right)}}\right], 
\end{eqnarray}
%
%%%%%%%%%%%%%%%%%%%%%%%%%%%%%%%%%%%%%%%%%%%%%%%%%%%%%%%%%%
%
where the symbols $N\left(\omega\right)$ and $f\left(\omega\right)$ denote the functions of Bose-Einstein and Fermi-Dirac, respectively.

The Eliashberg equations on the real axis have been solved using the numerical methods presented in the papers \cite{Radek06}, \cite{Radek07}. The convergent solutions have been obtained in the identical temperature range as for the Eliashberg equations on the imaginary axis.

%%%%%%%%%%%%%%%%%%%%%%%%%%%%%%%%%%%%%%%%%%%%%%%%%%%
\subsection{The energy gap}
%%%%%%%%%%%%%%%%%%%%%%%%%%%%%%%%%%%%%%%%%%%%%%%%%%%

%
%%%%%%%%%%%%%%%%%%%%%%%%%%%(Rysunek.4.)
\begin{figure*}[ht]
\includegraphics[scale=0.25]{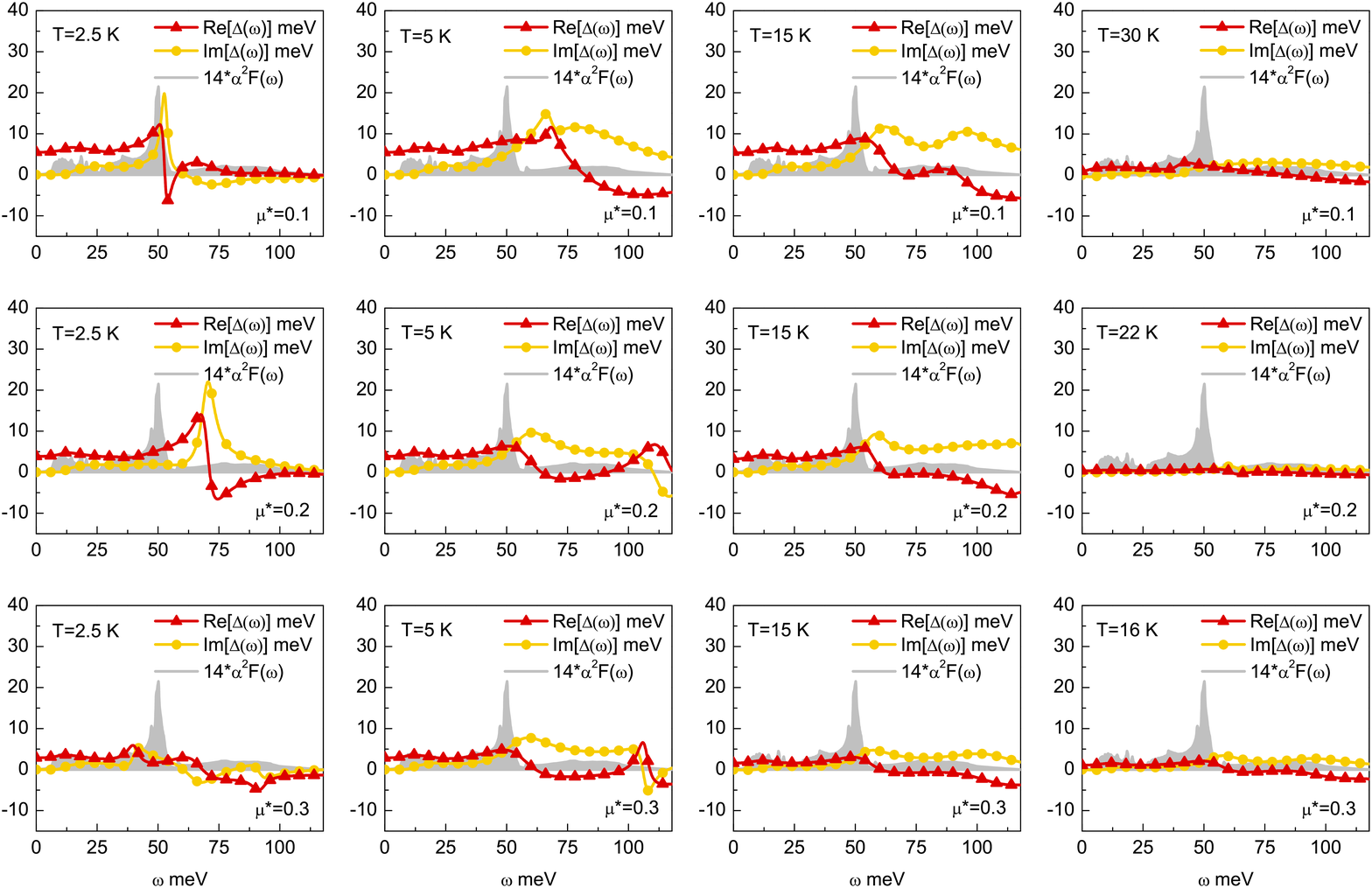}
\caption{The dependence of the order parameter on the frequency for the selected values of the temperature and the Coulomb pseudopotential. Additionally, the form of the rescaled Eliashberg function has been plotted.}
\label{f4}
\end{figure*}
%%%%%%%%%%%%%%%%%%%%%%%%%%%
%

Figure \ref{f4} presents the order parameter form on the real axis. The frequency range, at which the Eliashberg function is defined has been chosen ($\omega\in\left<0,\Omega_{\rm max}\right>$). We have considered the selected values of the temperature and the selected values of the Coulomb pseudopotential.

It is easy to notice that the order parameter takes the complex values. The real part of the function $\Delta\left(\omega\right)$ is used to determine the energy gap at the Fermi level, whereas the imaginary part of the order parameter determines the damping effects \cite{Varelogiannis}. In the case of low frequencies, and thus the physical area of the energy gap, the following occurs: ${\rm Re}\left[\Delta\left(\omega\right)\right]\gg{\rm Im}\left[\Delta\left(\omega\right)\right]$, which means small damping effects. For higher frequencies, the course of the real and imaginary parts of the order parameter is very complicated. However, for $T=2.5$ K and $\mu^{\star}=0.1$, the visible correlation between the shape of the order parameter and the Eliashberg function can be clearly seen. As it turns out, this correlation disappears with the increasing value of the temperature and the parameter $\mu^{\star}$. For the frequencies much higher than $\Omega_{\max}$, the order parameter becomes the subject of the saturation.

The energy gap at the Fermi level ($2\Delta$) can be calculated using the equation: 
\begin{equation}
\label{r13}
\Delta\left(T\right)={\rm Re}\left[\Delta\left(\omega=\Delta\left(T\right)\right)\right].
\end{equation}

From the physical point of view, the most important is the value of the energy gap at the lowest temperature ($T_{0}$). In the considered case, the following result has been obtained: $2\Delta\left(0\right)\in\left<11.21,5.78\right>$ meV. Hence, the dimensionless ratio 
$R_{\Delta}\equiv 2\Delta\left(0\right)/k_{B}T_{C}$ changes in the range of the values from $4.25$ to $3.98$. It should be noted that the BCS theory provides the universal value of $R_{\Delta}$ for all the superconductors: $\left[R_{\Delta}\right]_{\rm BCS}=3.53$ \cite{BCS}. The result obtained for $\left[R_{\Delta}\right]_{\rm PtH}$ confirms the fact that the properties of the superconducting state, can not be properly determined in the framework of the BCS model.

%%%%%%%%%%%%%%%%%%%%%%%%%%%%%%%%%%%%%%%%%%%%%%%%%%%
\section{Summary}
%%%%%%%%%%%%%%%%%%%%%%%%%%%%%%%%%%%%%%%%%%%%%%%%%%%

The study has characterized the thermodynamic properties of the superconducting state in the PtH compound under the pressure at $76$ GPa. Due to the high value of the electron-phonon coupling constant, the calculations have been carried in the framework of the Eliashberg formalism. A wide range of the Coulomb pseudopotential values has been chosen: $\mu^{\star}\in\left<0.1,0.3\right>$.

It has been found that the critical temperature is relatively high in the whole range of the Coulomb pseudopotential: $T_{C}\in\left<30.6,16.8\right>$ K. The values of the remaining thermodynamic parameters differ significantly from the expectations of the BCS theory. In particular, the dimensionless ratios of the characteristic values of the thermodynamic functions are respectively equal to: $R_{H}\in\left<0.145,0.156\right>$, $R_{C}\in\left<2.07,1.97\right>$, and $R_{\Delta}\in\left<4.25,3.98\right>$.
%%%%%%%%%%%%%%%%%%%%%%%%%%%%%%%%%%%%%%%%%%%%%%%%%%%

%%%%%%%%%%%%%%%%%%%%%%%%%%%%%%%%%%%%%%%%%%%%%%%%%%%
\begin{acknowledgements}
D. Szcz{\c{e}}{\'s}niak would like to acknowledge the financial support within the "Young Scientists" program, provided by the Dean of the Faculty of Mathematics and Science JDU (grant no. DSM/WMP/11/2012/26/).
\end{acknowledgements}
%%%%%%%%%%%%%%%%%%%%%%%%%%%%%%%%%%%%%%%%%%%%%%%%%%%

%%%%%%%%%%%%%%%%%%%%%%%%%%%%%%%%%%%%%%%%%%%%%%%%%%%

%%%%%%%%%%%%%%%%%%%%%%%%%%%%%%%%%%%%%%%%%%%%%%%%%%%%%%%%%%%%%%%%%%%%%%%%%%%%%%%%%%%%%%%%%%%%%%%%%%%
%
\end{document}